\documentclass[preprint,amsmath,amssymb,superscriptaddress,aps]{revtex4}
\usepackage{graphicx}
\usepackage{dcolumn}
\usepackage{bm}
\usepackage{color}
\usepackage{pifont}

\begin{document}


\title{Thermal conductivity of strong coupling V$_{1-x}$Ti$_x$ superconductors in the Mott-Ioffe-Regel limit \\}

\author{Sabyasachi Paul}{}
  \affiliation{FEL Utilization Laboratory, Raja Ramanna Center for Advanced Technology, Indore, Madhya Pradesh - 452 013, India}
  \affiliation{Homi Bhabha National Institute, Training School Complex, Anushakti Nagar, Mumbai 400 094, India}
 \author{L. S. Sharath Chandra}
 \email{lsschandra@rrcat.gov.in}
 \affiliation{FEL Utilization Laboratory, Raja Ramanna Center for Advanced Technology, Indore, Madhya Pradesh - 452 013, India}
\author{M. K. Chattopadhyay}
   \affiliation{FEL Utilization Laboratory, Raja Ramanna Center for Advanced Technology, Indore, Madhya Pradesh - 452 013, India}
   \affiliation{Homi Bhabha National Institute, Training School Complex, Anushakti Nagar, Mumbai 400 094, India}

\date{\today}

\begin{abstract}
We report an enhancement of thermal conductivity ($\kappa$) below the superconducting transition temperature ($T_C$) in the high carrier density $\beta$-V$_{1-x}$Ti$_x$ alloys. We find that the point defects generated when Ti is added to V reduce the electron mean free path down to the inter-atomic distances and make the high frequency phonons ineffective in carrying heat. In this Mott-Ioffe-Regel limit, the phonon thermal conductivity is dominated by the low frequency phonons limited by the scattering due to the electrons. The formation of Cooper pairs below the $T_C$ re-normalizes the phonon mean free path and enhances the $\kappa$.
\end{abstract}

\maketitle

The conductivity of a metal reduces when the disorder increases and it becomes an insulator at an extreme level of disorder \cite{mot90}. This metal to insulator transition occurs due to the localization of electrons when the electron mean free path ($l_e$) reduces to the inter-atomic distance $a$. This limit is called the Mott-Ioffe-Regel limit, and the transition is called the Anderson  transition \cite{mot90}. Although superconductivity around Anderson transition has attracted researchers for several decades \cite{and59, kec76, bel87, bel94, bur12, bur15}, the understanding on the nature of the superconductivity is yet to be clear \cite{bel94, bur12, bur15}. These studies became intense after the discovery of high temperature superconductors \cite{bed86} and the Fe based superconductors \cite{tes09, yan09} due to the underlined superconductor to insulator transition (in 2-dimensional materials) as a function of disorder and magnetic field which is thought to be a quantum phase transition \cite{por17, bre17, per17, fei18}. The loss of electron degrees of freedom in metals near the Anderson transition can also lead to a renormalization of electron-phonon coupling and an unconventional superconductivity \cite{kec76, rei86, bel87}. Such a renormalization results in the increase of the thermal conductivity ($\kappa$) when the temperature is reduced below the superconducting transition temperature ($T_C$) in the high temperature superconductors \cite{tew89, wil93, cas97} as well as in the amorphous superconductors \cite{gra77, loh80, esq83, gro86, loh91}, where the heat transfer in the normal state of these materials is mainly by the phonons (as in the case of insulators) \cite{loh81}. However, even in the disordered crystalline 3$d$ and 4$d$ electron superconductors, the heat is carried mainly by electrons and the $\kappa$ decreases when the temperature is reduced below the $T_C$ due to the reduction in the normal electron density \cite{bar59, sou69}. 

The thermal conductivity of metals can be expressed as $\kappa$ = $\kappa_e$ +$\kappa_l$, where $\kappa_e$ = $\frac{1}{3}C_e v_F l_e$ and $\kappa_l$ = $\frac{1}{3}C_l v_s l_{ph}$ are the electronic and lattice thermal conductivities respectively \cite{tri04}. Here, $C_e$,  and $C_l$ are the electronic and lattice heat capacities respectively. The $v_F$ is the Fermi velocity, $v_s$ is the sound velocity, and $l_{ph}$ is the phonon mean free path. For a 3$d$ transition metal, it turns out that the ratio $\kappa_l/\kappa_e \approx$ 10$^{-5}$T$^2 l_{ph}/l_e$. This indicates that the electrons are the major carriers of heat at low temperatures in these materials. The $\kappa_e$ can become comparable to the $\kappa_l$ only when the $l_e$ reduces to inter atomic distances in a crystalline medium. Achieving $\kappa_l > \kappa_e$ even in the disordered crystalline metals is nearly impossible. However, we find that in certain disordered crystalline V$_{1-x}$Ti$_x$ alloys in the Mott-Ioffe-Regel limit, such a condition is achieved due to the loss of normal electrons when these alloys become superconducting.

The body centred cubic (bcc) $\beta$-V$_{1-x}$Ti$_x$ alloys are considered to be an alternate system for superconducting magnet applications due to their relatively high $T_C$ and large upper critical field in the zero temperature limit ($H_{C2}$($T = 0$)) \cite{mat14}. They are also promising superconducting materials for the neutron radiation environment \cite{tai07, mat14}. The $T_C$ of these alloys are influenced by strong electron-phonon interaction as well as spin fluctuations \cite{mat14, mat14a}. Although, these alloys are mechanically strong, their dissipationless current carrying capacity is nearly two orders less than the commercial Nb-Ti alloys \cite{mat13, mat15}. Controlled introduction of defects might improve the current carrying capability, while at the same time degrading the thermal conductivity and thus the ability of the material to remove the dissipated heat. Here, we show, however, that with decreasing temperature below the $T_C$, the thermal conductivity of certain V$_{1-x}$Ti$_x$ superconducting alloys  increases four times the value at the $T_C$ as a result of the re-normalization of the phonon mean free path. 

\begin{figure}
\includegraphics[width = 8.5cm,height = 13cm]{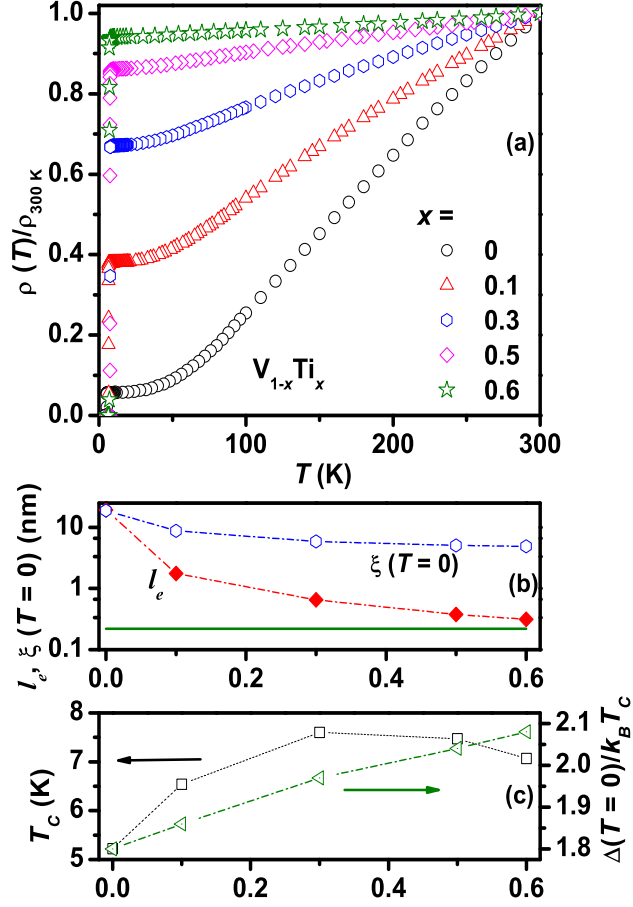}
\caption{\label{fig:epsart} (color online)  (a) Temperature dependence of resistivity of the V$_{1-x}$Ti$_x$ alloys normalized at 300~K. As $x$ increases, the  $\rho$(8~K) increases and for $x$ = 0.6, the resistivity become nearly independent of temperature. (b) The $l_e$ (solid symbols), the $\xi$($T = 0$) (open symbols) and the $a$ (green solid line) for the above alloys indicate that $l_e$ approaches $a$ as $x$ is increased, while the $\xi$($T = 0$) is more than an order  higher than $l_e$. (c) Both $T_C$ as well as $\Delta$($T = 0$)$/k_BT_C$ increases with $x$, however, the $T_C$ reduces at higher concentrations of titanium.} 
\end{figure}

Figure 1(a) shows the temperature dependence of resistivity ($\rho$($T$)) of the V$_{1-x}$Ti$_x$ alloys plotted as $\rho$($T$)/$\rho$(300~K). The residual resistivity $\rho$(8~K) = 1.29~$\mu \Omega$ cm of vanadium is about 18 times smaller than $\rho$(300~K). The $\rho$(8~K) increases to more than 90~$\mu \Omega$ cm when the titanium concentration in vanadium increases to 60 at.\% and beyond. These results are consistent with our previous studies \cite{mat14}. The $\rho$($T$) is nearly independent of temperature for the $x$ = 0.6 alloy, whereas for the $x$ = 0.7 alloy, a negative temperature coefficient of resistivity is observed at low temperatures. Figure 1(b) shows the $l_e$ estimated from $\rho$(8~K) and the superconducting coherence length $\xi$($T = 0$) estimated from the H$_{C2}$(T=0).  The $l_e$ is comparable to the $a$ = 0.22~nm (green solid line) for the alloys with $x \geq$ 0.1 whereas $\xi$($T = 0$) is about an order of magnitude higher than the $l_e$. These results indicate that these alloy superconductors are in the vicinity of the Mott-Ioffe-Regel limit. The Fig. 1(c) shows the $T_C$ and the normalized superconducting energy gap $\Delta(T=0)/k_BT_C$ ($k_B$ is the Boltzmann constant) estimated from the temperature dependence of heat capacity. The $\Delta(T=0)/k_BT_C$ is larger than the Bardeen-Cooper-Schrieffer (BCS) limit of 1.76. 

Figure 2 shows the temperature dependence of the thermal conductivity in the range 2-8~K in the superconducting ($\kappa_s$) and normal states ($\kappa_n$) of the V$_{1-x}$Ti$_x$ alloys. The $\kappa_s$ is measured in the absence of applied magnetic field and $\kappa_n$ is obtained by performing the measurements in 8~T magnetic field. The $\kappa_n(T)$ of the alloys is an order of magnitude less than that of vanadium. A gross estimation of electronic part of thermal conductivity ($\kappa_{en}$) from the Wiedermann-Franz law ($\kappa_{en} = L_0T/\rho(8~K)$, $L_0$ being Lorenz number) shows that the heat is mainly carried by the electrons in the normal state of all these alloys. The reduction in the normal electrons below $T_C$ results in the reduction of the $\kappa_s$ in comparison with the $\kappa_n$ of vanadium. In contrast, there is only a small difference between $\kappa_n$ and $\kappa_s$ of the $x$ = 0.1 alloy.  In all the other alloys with $x >$ 0.1, the $\kappa_s > \kappa_n$ for $T < T_C$, indicating that the phonons are the major carriers of heat in the superconducting state. In the literature, larger $\kappa_s$ in comparison to $\kappa_n$ is seen in systems like the high temperature ceramic superconductors ($T_C \sim$ 100~K) \cite{ric92, cas97}, ceramic NbC ($T_C \sim$ 10~K) \cite{rad69}, and amorphous superconductors such as Zr$_{70}$Cu$_{30}$ ($T_C \sim$ 2.7~K) \cite{loh81} where $\kappa_{en}$ is only about 20\% of the total thermal conductivity. The free carrier density of ceramic superconductors are about two orders less than that of the good metals like Copper \cite{wil88, wil93}. In the amorphous superconductors, there is a lack of long range order. Therefore, it is natural that the phonons are the major carriers of heat in these systems. However, the present crystalline V$_{1-x}$Ti$_x$ alloys have high free electron density. Therefore, it is rather surprising to have $\kappa_s > \kappa_n$ in spite of electrons being the major carriers of heat in the normal state. To understand this, we have separated the electronic ($\kappa_{ez}$) and lattice ($\kappa_{lz}$) contributions to the $\kappa(T)$ in both the normal ($z = n$) and the superconducting ($z = s$) states.          

\begin{figure*}

\includegraphics[width = 12cm,height = 16cm]{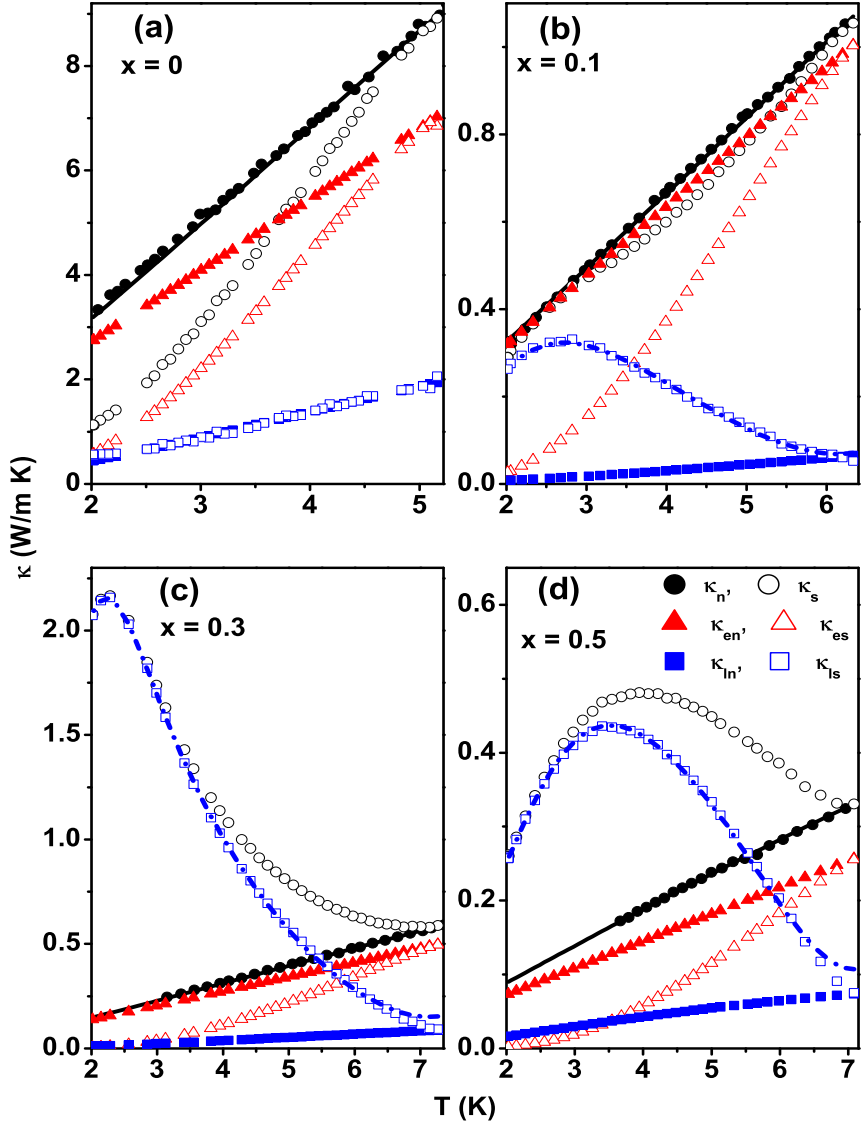}
\caption{\label{fig:wide}(color online) Temperature dependence of $\kappa$ in the V$_{1-x}$Ti$_x$ alloys below the $T_C$ in the superconducting (zero magnetic field, open symbols) and normal states (8~T, closed symbols). The $\circ$ and $\bullet$ represent the experimental data points, {\color{red} $\vartriangle$} and {\color{red} $\blacktriangle$} represent the $\kappa_e$, {\color{blue} $\square$} and {\color{blue}$\blacksquare$} represent the $\kappa_l$. The solid (black) line is the fit to the $\kappa_n$ and  the dash-dotted (blue) line is the fit to the $\kappa_{ls}$. In the superconducting state of vanadium, the heat is carried mainly by the electrons. The $\kappa$($T$) in the superconducting state of the V$_{1-x}$Ti$_x$ alloys with $x >$ 0.1 increases when the temperature is decreased below $T_C$ indicating that the phonons are the major carriers of heat in the superconducting state.}
\end{figure*}

The $\kappa_{ez}$ is given by  $\kappa_{ez}^{-1} = \kappa_{e-i,z}^{-1}+\kappa_{e-l,z}^{-1} $, where $\kappa_{e-i,z}^{-1}$ and $\kappa_{e-l,z}^{-1}$ are the thermal resistivities due to the scattering of electrons from the defects and phonons respectively. In the normal state, $\kappa_{e-i,n}^{-1} = AT^{-1}$ and $\kappa_{e-l,n}^{-1} = BT^2+\mathcal{O}(T^ 4)$, where $A$ and $B$ are the constants. The exact form of  $\kappa_{e-l,n}$ can be found in ref. \cite{tri04}. In the superconducting state, $\kappa_{e-i,s} = \kappa_{e-i,s-n} \times \kappa_{e-i,n} $ and $\kappa_{e-l,s} = \kappa_{e-l,s-n} \times \kappa_{e-l,n}$, where the ratios $\kappa_{e-i,s-n} =\kappa_{e-i,s}/\kappa_{e-i,n}$  and $\kappa_{e-l,s-n} = \kappa_{e-l,s}/\kappa_{e-l,n}$ are given by Bardeen et. al., \cite{bar59}. 

The $\kappa_{lz}$ is given by \cite{tew89}

\begin{equation}
\kappa_{lz} =  MT^3 \int^{\infty}_{0} dx x^4 e^x (e^x -1)^{-2}\tau,
\end{equation}

where $\tau^{-1} = N_{z} + L_{z} x T + C_{z}J_{3}^{-1}(\theta_D/T) g(x) x T + P_{z} x^4 T^4$ and the constant $M = k_B^4 /(2\pi^2 \hbar^3 v_s)$. Here, $\hbar$ is the reduced Plank's constant, $v_s$ is the sound velocity. The coefficients $N_{z}$, $L_{z}$, $C_{z}$ and $P_{z}$ represent the strength of phonon scattering due to boundaries, dislocations, electrons and point defects respectively. The $J_{3}(\theta_D/T)$ is the Debye function \cite{tri04}, $g(x)$ is the ratio of phonon scattering by electrons in the superconducting and normal states \cite{bar59} and $x$ is the reduced phonon energy ($x = h\omega/k_BT$, $\omega$ is the phonon frequency). The difference between $\kappa_{ln}$ and $\kappa_{ls}$ is that the $g(x)$ is equal to unity in the normal state and the form of $g(x)$ in the superconducting state is in Ref. \cite{bar59}.
 
The symbols {\color{red} $\blacktriangle$ } and {\color{blue} $\blacksquare$ } in Fig.2., represent the $\kappa_{en}$ and $\kappa_{ln}$  respectively. The $\kappa_{es}$ ({\color{red} $\vartriangle$}) is obtained by the inverting the sum of  $\kappa_{e-i,s}^{-1}$ and $\kappa_{e-l,s}^{-1}$. The  $\kappa_{e-i,s}$ and $\kappa_{e-l,s}$ are estimated using experimentally obtained $\Delta(T = 0)/k_BT_C$. The $\kappa_{ls}$ ({\color{blue} $\square$}) is obtained by subtracting $\kappa_{es}$ from the $\kappa_{s}$. The $\kappa_s$ of vanadium resembles that of niobium \cite{sha12} where $\kappa_s$ lies below the $\kappa_n$ for $T_C/2 < T < T_C$. The temperature dependence of $\kappa_s$ is due to the temperature dependence of $\kappa_{es}$ as there is no difference between $\kappa_{ln}$ and $\kappa_{ls}$ in the range  $T_C/2 < T < T_C$. This is a perfect example for the case where the electrons dominate the heat conduction. The analysis of the $\kappa_n$ of the alloys indicates that the $\kappa_{en}$ contributes more than 70\% to the $\kappa_n$ below 8~K, although the $l_e$ approaches the Mott-Ioffe-Regel limit. This indicates that these alloys are at the metallic side of the Anderson transition \cite{mot87}. 

We found by fitting $\kappa_{ls}$ using eq.1 that the thermal conduction by phonons in the normal state of these alloys is limited only by scattering due to the electrons and point defects. In case of vanadium,  $\kappa_{es} > \kappa_{ls}$ for $T \geq$ 2~K.  However, the $\kappa_{ls}$ of the $x = 0.1$ alloy starts to increase when the temperature is reduced below $T_C$ = 6.4~K and become more than $\kappa_{es}$ below a temperature T$_{cr}$ = 3.4~K which is more than $T_C/2$. In case of $x$ = 0.3 and 0.5 alloys, the $T_{cr}$  is very close to $T_C$. Similar effect is observed in the alloys with $x$ up to 0.7 (not shown here) above which the bcc structure become unstable and a transition to a hexagonally close packed structure \cite{col86} is observed at low temperatures thereby avoiding a metal-insulator transition. Since, the mean free path of the phonons scattered by point defects varies as $l_{ph}^{-1} \propto \omega^4$, the high frequency phonons are disproportionately scattered, leaving a large part of the heat conduction to long wavelength phonons \cite{ric92}. The average mean free path of the phonons ($l_{ph}$) obtained from $\kappa_{ln}$ of these alloys is about 30-100~nm, where we have estimated $v_s$ ( = 2650~m/s) from the elastic constants \cite{col86}. The metallography results (not shown here) show that the grain size of the present alloys ranges from few tens of microns to few millimeters. Therefore, the scattering of phonons from the grain boundaries is negligible. Metallography results further indicate that the dislocation density is also small. The electron-phonon interaction in these V$_{1-x}$Ti$_x$ alloys being quite strong  \cite{mat14}, the low frequency phonons are scattered by electrons alone in the normal state. In the superconducting state, the loss of normal electrons makes these phonons much more effective in carrying heat as the $l_{ph}$ increases in comparison with that of the normal state. Therefore, the scattering of phonons by the point defects becomes less effective. Instead, dislocations become the major scatterers of phonons and $\kappa_{ls}$ decreases when temperature is reduces below $T_C/2$.         

\begin{figure}
\includegraphics[width = 8.5cm,height = 13cm]{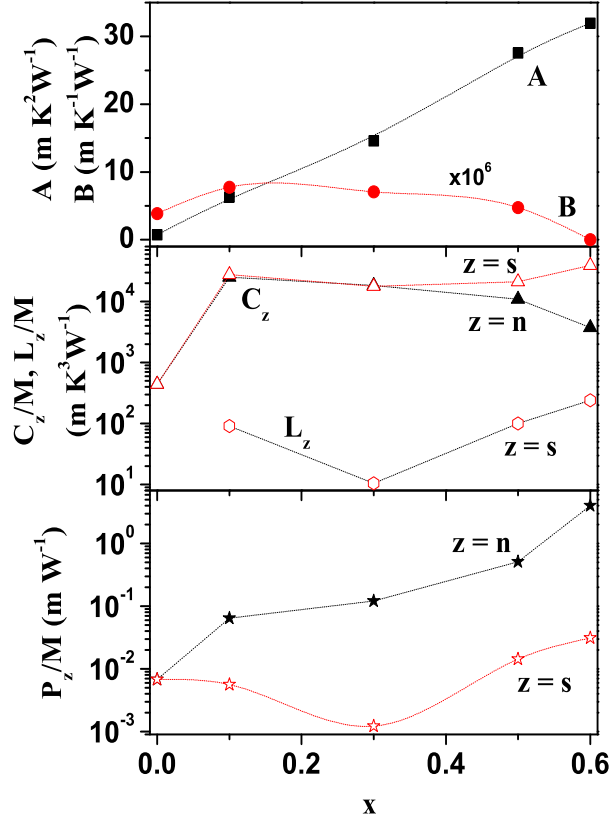}
\caption{\label{fig:epsart} (color online)  Coefficients of the temperature dependence of thermal resistivity due to various scattering mechanisms: A and B are coefficients for the temperature dependence of thermal resistivity in the normal state due to  the scattering of electrons by the defects and the phonons respectively. The $C_z$, $L_z$ and $P_z$ are respectively for the thermal resistance faced by phonons due to the scattering by electrons, dislocations and point defects. The subscripts $z$ = n and s stand for the normal state and superconducting state respectively. }
\end{figure}

Figure 3 shows the variation of different coefficients corresponding to the different scattering events that limit the thermal conductivity in the V$_{1-x}$Ti$_x$ alloys. Figure 3(a) shows that the coefficients $A$ and $B$ of the thermal resistivity due to the scattering of electrons from the defects and phonons respectively. It is observed that the contribution from the  scattering of elections from the phonons is negligible in comparison to that from the scattering of electrons from the defects ($B << A$). The $A$ increases linearly with the composition. Majority of the these defects are the point defects generated by the addition of titanium. As the amount of titanium increases in vanadium, the number of point defects increases and the resistance to heat flow increases. However, strong electron-phonon interaction persists all through the composition range due to the softening of the phonons when titanium is added to vanadium \cite{mat14}. Therefore, the coefficient $C_n$  increases about two orders of magnitude when 10 at.\% of Ti is added to vanadium (Fig. 3(b)). The $C_n$ decreases for large values of $x$ due to the reduction in the electron-phonon interaction and due to the incipient instability of the lattice. However, $C_s$ remains almost constant (Fig. 3(b)) for all the alloys studied here as thermal conduction in the superconducting state of the V$_{1-x}$Ti$_x$ alloys is hardly limited by the electron-phonon scattering. Fig. 3(c) shows that the coefficient $P_n$ increases with $x$ due to the increase in the point defects. However, $P_s$ is an order less than $P_n$ due to the re-normalization of $l_{ph}$ when these alloys become superconducting. Therefore, the dislocations become effective scattering centres (Fig. 3(b)) in the superconducting state at low temperatures ($T < T_C/2$).
 
 We conclude that the observation of an increase in the thermal conductivity when the temperature is reduced below $T_C$  in a crystalline superconductor having dense free electrons requires the presence of a very large number of point defects. The point defects scatter electrons and phonons differently. When present in very large numbers, the point defects reduce $l_e$ drastically to the inter-atomic distances driving the material towards metal-insulator transition, while they scatter effectively the high frequency phonons leaving out the heat conduction to long wavelength phonons. In such a scenario, if the electron-phonon scattering is strong, these long wave length phonons are scattered mainly by the electrons as the average $l_{ph}$ is quite small as compared to the sample dimensions, and the inter-grain boundary and inter-dislocations distances. In such a case, the loss of electrons when the material become superconducting enables the phonons to carry heat effectively. This results in the enhanced thermal conductivity below the $T_C$ which in general is not expected in a metallic system.          
 
{\label{1} { \fontsize{8}{0.2} \it Methods: The poly-crystalline V$_{1-x}$Ti$_x$ samples were prepared by arc melting and characterized by x-ray diffraction measurements (see the supplement for the details). The electrical resistivity and thermal conductivity measurements on these alloys were performed using a physical property measurement system (Quantum Design, USA).} }

{\it Acknowledgements:} We thank Dr. S. B. Roy for helpful discussions.



\begin{thebibliography}{}

\bibitem{mot90} See eg. N. F. Mott, {\it Metal-Insulator Transitions} (Taylor and Francis, London (1990)) 
\bibitem{and59} P. W. Anderson, J. Phys. Chem. Solids {\bf 11}, 26 (1959).
\bibitem{kec76} B. Keck, and A. Schmid, J. Low Temp. Phys. {\bf 24}, 611 (1976).
\bibitem{bel87} D. Belitz, Phys. Rev. B {\bf 36}, 47 (1987).
\bibitem{bel94} D. Belitz, and T. R. Kirkpatrick, Rev. Mod. Phys. {\bf 66}, 261 (1994) and the references therein.
\bibitem{bur12} I. S. Burmistrov, I. V. Gornyi, and A. D. Mirlin, Phys. Rev. Lett. {\bf 108}, 017002 (2012).
\bibitem{bur15} I. S. Burmistrov, I. V. Gornyi, and A. D. Mirlin, Phys. Rev. B {\bf 92}, 014506 (2015). 
\bibitem{bed86} J. G. Bednorz, and K. A. Muller, Z. Phys. B {\bf 64}, 189 (1986).
\bibitem{tes09} Z. Tesanovic, Phys. {\bf 2}, 60 (2009).
\bibitem{yan09} W. L. Yang, A. P. Sorini, C-C. Chen, B. Moritz, W.-S. Lee, F. Vernay, P. Olalde-Velasco, J. D. Denlinger, B. Delley, J.-H. Chu, J. G. Analytis, I. R. Fisher, Z. A. Ren, J. Yang, W. Lu, Z. X. Zhao, J. van den Brink, Z. Hussain, Z.-X. Shen, and T. P. Devereaux, Phys. Rev. B {\bf 80}, 014508 (2009) and the references therein.
\bibitem{por17} S. Poran, T. Nguyen-Duc, A. Auerbach, N. Dupuis, A. Frydman, and O. Bourgeois, Nat. Commun. {\bf 8} 14464 (2017).  
\bibitem{bre17} N. P. Breznay, M. Tendulkar, L. Zhang. S-C. Lee, and A. Kapitulnik, Phys. Rev. B {\bf 96} 134522 (2017).
\bibitem{per17} I. M. Percher, I. Volotsenko, A. Frydman, B. I. Shklovskii, and  A. M. Goldman, Phys. Rev. B {\bf 96} 224511 (2017).
\bibitem{fei18} M. V. Feigel'man and L. B. Ioffe, Phys. Rev. Lett. {\bf 120} 037004 (2018)
\bibitem{rei86} M. Yu. Reizer, and A. V. Sergeyev, Sov. Phys. JETP {\bf 63} 616 (1986) [Zh. Eksp. Teor. Fiz. {\bf 90}, 1056 (1986)].
\bibitem{tew89} L. Tewordt, and Th. W{\"o}lkhausen, Solid State Commun. {\bf 70}, 839 (1989) and the references therein.
\bibitem{wil93} W. S. Williams, Solid State Commun. {\bf 87}, 355 (1993).
\bibitem{ric92} R. A. Richardson, S. D. Peacor, C. Uher, and F. Nori, J. Appl. Phys. {\bf 72} 4788 (1992).
\bibitem{cas97} S. Castellazzi, M. R. Cimberle, C. Ferdeghini, E. Giannini, G. Grasso, D. Marr{\`e}, M. Putti, A. S. Siri, Physica C {\bf 273}, 314 (1997).
\bibitem{gra77} J. E. Graebner, B. Golding, R. J. Schutz, F. S. L. Hsu, and H. S. Chen, Phys. Rev. Lett. {\bf 39}, 1480 (1977).
\bibitem{loh80} H. v. L{\"o}hneysen, M. Platte, W. Sander, H. J. Schink, G. v. Minnigerode, and R. Samwer, J. Phys. C Colloq. {\bf 8}, 745 (1980).
\bibitem{esq83} P. Esquinazi, and F. de la Cruz, Phys. Rev. B {\bf 27}, 3069 (1983).
\bibitem{gro86} H. W. Gronert, D. M. Herlach, A. Schr{\"o}der, R. van den Berg, and H. v. L{\"o}hneysen, Z. Phys. B {\bf 63}, 173 (1986).
\bibitem{loh91} H. v. L{\"o}hneysen, Mat. Sci. Eng. A {\bf 133} 51 (1991).
\bibitem{loh81} H. v. L{\"o}hneysen, D. M. Herlach, E. F. Wassermann, and K. Samwer, Solid State Commun. {\bf 39}, 591 (1981).
\bibitem{bar59} J. Bardeen, G. Rickayzen, and L. Tewordt, Phys. Rev. {\bf 113}, 982 (1959).
\bibitem{sou69} J. B. Sousa, J. Phys. C: Solid State Phys. {\bf 2}, 629 (1969).
\bibitem{tri04} See e. g. T. M. Tritt, {\it Thermal Conductivity-Theory, Properties and Applications} (Kluwer Academic/Plenum Publishers, New York (2004))
\bibitem{tai07} M. Tai, K. Inoue, A. Kikuchi, T. Takeuchi, T. Kiyoshi, Y. Hishinuma, IEEE Trans. Appl. Supercond. {\bf 17}, 2542 (2007) (and references therein).
\bibitem{mat14} Md. Matin, L. S. Sharath Chandra, S. K. Pandey, M. K. Chattopadhyay, and S. B. Roy, Eur. Phys. J. B {\bf 87}, 131 (2014).
\bibitem{mat13} Md. Matin, L. S. Sharath Chandra, M. K. Chattopadhyay, R. K. Meena, Rakesh Kaul, M. N. Singh, A. K. Sinha,  and S. B. Roy, J. Appl. Phys. {\bf 113}, 163903 (2013).
\bibitem{mat15}Md. Matin, L. S. Sharath Chandra, M. K. Chattopadhyay, R. K. Meena, Rakesh Kaul, M. N. Singh, A. K. Sinha, S. B. Roy, Physica C {\bf 512},  32 (2015). 
\bibitem{mat14a} Md. Matin, L. S. Sharath Chandra, R. K. Meena, M. K. Chattopadhyay, A. K. Sinha, M. N. Singh and S. B. Roy, Physica B {\bf 436}, 20 (2014).
\bibitem{rad69} L. G. Radosevich, and W. S. Williams, Phys. Rev. {\bf 188} 770 (1969).
\bibitem{wil88} W. S. Williams, Mat. Sci. Engg. A {\bf 105/106} 1 (1988).
\bibitem{mot87} N. Mott, J. Phys. C: Solid State Phys. {\bf 20} 3075 (1987).
\bibitem{sha12} L. S. Sharath Chandra, M. K. Chattopadhyay, S. B. Roy, V. C. Sahni, and G. R. Myneni, Supercond. Sci. Technol. {\bf 25} 035010 (2012).
\bibitem{col86}  E. W. Collings, {\it Applied Superconductivity, Metallurgy, and Physics of Titanium Alloys} {\bf 1 \& 2}, (Plenum Press, NewYork (1986)).





\end{thebibliography}

\end{document}